\documentstyle[twoside,fleqn,espcrc2,epsf]{article}

\newcommand{\AmS}{{\protect\the\textfont2
  A\kern-.1667em\lower.5ex\hbox{M}\kern-.125emS}}

\title{ 
       \vspace{-3.65cm}                                     
       {\normalsize DESY 96--152}    \\[-0.2cm]             
       {\normalsize HUB--EP--96/37} \\[-0.2cm]              
       {\normalsize August 1996}   \\                       
       \vspace{2.25cm}                                      
 The Drell-Yan process and Deep Inelastic Scattering from the lattice%
        \thanks{Talk presented by P. Rakow at Lat96, St. Louis, U.S.A.}}

\author{M.~G\"ockeler%
           \address{H{\"o}chstleistungsrechenzentrum HLRZ,
                    c/o Forschungszentrum J{\"u}lich, D-52425 J{\"u}lich,
                                                             Germany}
           \hspace{-0.2cm}
           \address{Institut f\"ur Theoretische Physik, RWTH Aachen,
                    D-52056 Aachen, Germany}
           \hspace{-0.2cm}
           \address{Institut f\"ur Theoretische Physik, J.~W. Goethe
                    Universit\"at, D-60054 Frankfurt, Germany},
        R.~Horsley%
           \address{Institut f\"ur Physik, Humboldt-Universit\"at zu Berlin,
                    Invalidenstra{\ss}e 110, D-10115 Berlin, Germany},
        E.-M.~Ilgenfritz$^{\rm d}$,
        H.~Oelrich$^{\rm a}$
           \hspace{-0.2cm}
           \address{DESY-IfH Zeuthen, D-15735 Zeuthen, Germany},
        H.~Perlt%
           \address{Fak. f. Physik und Geowiss., Universit\"at Leipzig,
                    Augustusplatz 10--11, D-04109 Leipzig, Germany},
        P.~E.~L.~Rakow$^{\rm a}$ \hspace{-0.2cm} $^{\rm e}$,
        G.~Schierholz$^{\rm a}$ \hspace{-0.2cm} $^{\rm e}$
           \hspace{-0.2cm}
           \address{Deutsches Elektronen-Synchrotron DESY,
                    Notkestra{\ss}e 85, D-22603 Hamburg, Germany},
        A.~Schiller$^{\rm f}$
        and 
        P.~Stephenson$^{\rm e}$ }
       
\begin{document}

\begin{abstract}
  We report on measurements of the $h_1$
 structure function, relevant to calculating cross-sections for the Drell-
 Yan process.  This is a quantity which can not be measured in Deep Inelastic 
 Scattering, it gives additional information on the spin carried by the
 valence quarks, as well as insights on how relativistic the quarks are. 
\end{abstract}

\maketitle

\section{INTRODUCTION}

  Deep Inelastic Scattering (DIS) gave us the first evidence that
 quarks are true physical objects, 
 not just ``book-keeping'' devices for a flavour symmetry group. 
 The structure functions measured in DIS give us an important
 insight into the internal
 workings of a hadron, allowing us to measure how the
 energy and spin is shared out between the different constituents.

  Perturbative QCD can explain the evolution of these structure functions
 as we change the scale at which we probe the hadron, but is unable
 to give us a starting value for this evolution. Lattice QCD has been
 reasonably 
 successful for masses, the next step is to use 
 it to calculate structure function moments, form factors and 
 matrix elements. At present this is our only hope of finding these
 numbers from first principles. 

\section{THE INTERPRETATION OF $h_1$}

   Deep Inelastic Scattering (DIS) is not the only useful probe
 of the hadrons' parton distributions. 
In hadron-hadron colliders the 
 Drell-Yan process can be observed, in which a quark in one
 hadron and an anti-quark in the other annihilate to form an 
 extremely virtual time-like photon, which then decays to a 
 lepton-antilepton pair. Measurements of the lepton pair's 
 total momentum give enough information to extract $x$ and $y$, 
 the fraction each parton carries of its hadron's momentum.  

    If we look at the unpolarised Drell-Yan process we find the
 same structure functions occurring as in DIS. For example the 
 total cross-section is proportional to 
 $ \sum_a e_a^2 f_1^a(x) f_1^{\bar{a}}(y)$ where $a$ runs over all
 flavours.  The asymmetry in cross-sections when two 
 longitudinally polarised nucleons collide can again be expressed in terms 
 of a quantity known from polarised DIS, namely the structure function 
 $g_1$. However if we consider the cross-section   
 when two transversely
 polarised nucleons with spins $S_A$ and $S_B$ collide, and try to
 find the asymmetry on flipping one spin,
\begin{eqnarray}
 A_{TT} & \equiv &\frac
 {\sigma \left(S_A , S_B\right)  -\sigma \left(S_A , -S_B\right)} 
 {\sigma \left(S_A , S_B\right)  +\sigma \left(S_A , -S_B\right)}
 \nonumber \\
 & \propto & 
 \frac{ \sum_a e_a^2 h_1^a(x) h_1^{\bar{a}}(y) } 
       { \sum_a e_a^2 f_1^a(x) f_1^{\bar{a}}(y) }  \ \  , 
\end{eqnarray} 
\newpage \noindent
 we find that it can not be expressed in
 terms of the familiar structure functions, a new structure function, 
 $h_1$, is needed \cite{jaffe_h1}.  

\begin{figure}[hbt]
\vspace*{-0.4cm}
\hspace*{0.5cm} 
\epsfxsize=6.0cm \epsfbox{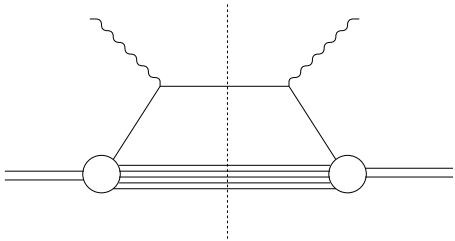}  
\vspace*{-0.7cm}
\caption{The Feynman diagram for deep inelastic scattering.
 The hard scattering process conserves chirality.}
\vspace*{-0.5cm}
\label{DISfeyn}
\end{figure}

 Moments of the structure function $h_1$ can be
 related, through the
 operator product expansion, to  the matrix elements of the
 operators 
\begin{equation}
 O^{\sigma , \mu_1 , \cdots , \mu_n }  \equiv
 \bar{\psi} i \sigma^{\sigma \left\{ \mu_1 \right. } \gamma_5 
 i D^{\mu_2} \cdots i D^{\left. \mu_n \right\} } \psi - \rm{tr.} 
 \label{h1_op} 
\end{equation} 
 This operator is very similar in structure to the operators that give
 the moments of $f_1$ and $g_1$. An important difference is in the 
 Dirac structure of the operator. In $f_1$ and $g_1$ this is proportional
 to $\gamma_\mu$ or $\gamma_\mu \gamma_5$ respectively, in both cases 
 these matrices anti-commute with $\gamma_5$. However the operator
 in Eq.~(\ref{h1_op}) is proportional to a $\sigma$ matrix, which 
 commutes with $\gamma_5$, implying that $h_1$ has the opposite 
 chiral properties to $f_1$ and $g_1$. It is this difference that 
 explains why $h_1$ is observable in the Drell-Yan process, but not
 in DIS. With massless quarks chirality is conserved along a quark
 line. In the Feynman diagram for DIS (Fig.~\ref{DISfeyn}) there is only
 a single quark line passing through the hard part of the process, so
 only the chirally-even structure functions $f_1$ and $g_1$ can be
 observed. On the other hand the Drell-Yan process involves two
 separate fermion lines, which can have the same or opposite chirality. 
 This means that, even at leading twist, both chirally-even and
 chirally-odd structure functions are observable. 
 
\begin{figure}[hbt]
\vspace*{0.6cm}
\epsfxsize=7.0cm \epsfbox{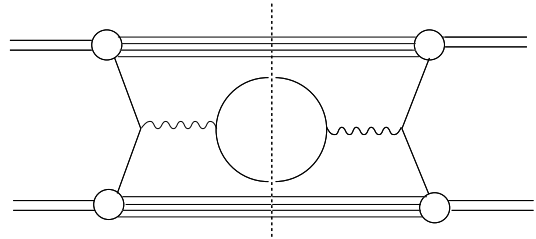}  
\vspace*{-0.4cm}
\caption{The Feynman diagram for the Drell-Yan process.
 Because two separate fermion lines are involved, chirality need not be
 conserved.}
\vspace*{-0.4cm}
\label{DYfeyn}
\end{figure}

  What does the structure function $h_1$ tell us about the quarks in
 the proton? If we consider a stationary proton (momentum in the 
 0-direction) the operator in Eq.~(\ref{h1_op}) differs from the 
 operator for $g_1$ by a factor $\gamma_0$. In the non-relativistic
 limit fermions are in eigenstates  of $\gamma_0$ with eigenvalue $1$, 
 and so $h_1$ and $g_1$ are identical. By comparing $h_1$ and $g_1$ for
 a real proton we can gain an insight into how relativistic the constituents
 are. 

  The structure functions $h_1$ and $g_1$ have the opposite
 behaviour under charge conjugation, so it is expected that the
 contributions of the quarks and anti-quarks in the sea will 
 largely cancel in $h_1$, making it a quantity given mostly by
 the valence quarks. This means that we might hope that a
 quenched calculation of $h_1$ is likely to give an answer close to
 the true value. 

   Beyond the non-relativistic approximation, what does $h_1$ measure?
 By considering the effect of operating with the operator from
 Eq.~(\ref{h1_op}) on
 a quark state which is an eigenstate of the operator 
 ${\not{\hspace*{-0.05cm}s}}_\perp \gamma_5$ \cite{jaffe_h1}
 it can be seen that  what is actually being measured
 by $h_1$ is the distribution of this quantity, which is given the 
 name ``transversity''. 

\section{RESULTS}

 The operator in Eq.~(\ref{h1_op}) can be discretised and its expectation 
 value measured on the lattice using the methods used earlier
 \cite{goeckeler95a}  
 for the more familiar $f_1$ and $g_1$ structure functions. 

 We have undertaken measurements of the quenched QCD structure functions. 
 The measurements I report here were made with the quenched
 Sheikholeslami-Wohlert
 action on a $16^3 \times 32$ lattice with $\beta = 6.0$ and 
 $c_{SW}=1.769$ (the value recommended in~\cite{magic_c}, chosen to
 eliminate $O(a)$ discretisation effects). To completely remove  
 all $O(a)$ effects it is not enough to simply use the optimal
 fermion action, the lattice operators must also be improved. At
 present we are using tree-level operator improvement. To produce
 a final answer we need to know the $Z$ factors of renormalisation 
 theory. We are using the 1-loop perturbative calculation of~\cite{perlt}. 
 Ideally we would like to determine both the operator improvement
 and the $Z$ factors non-perturbatively, we are currently working on
 this calculation. 

\begin{figure}[htb]
\vspace*{-1.8cm}
\hspace*{-1.2cm} 
\epsfxsize=9.5cm \epsfbox{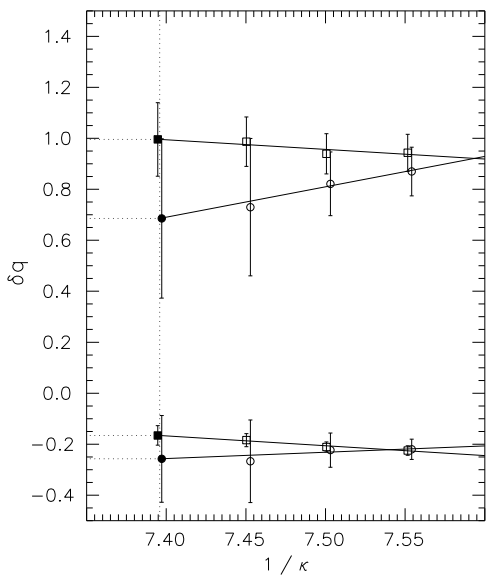}  
\vspace*{-4.2cm}
\caption{ The proton tensor charge 
 $\delta q \equiv \int_0^1 d x \, (h_1(x) -{\bar{h}}_1(x))$  
 for the valence $u$ (upper two lines) and $d$ (lower two lines) quarks. 
 The open points are the data, the solid points the chiral extrapolation.}  
\vspace*{-0.3cm}
\label{h1}
\end{figure}

  In Fig.~\ref{h1} we show our results for the proton  
 $\delta q \equiv \int_0^1 d x \, (h_1(x) - {\bar{h}}_1(x))$  
 for the $u$ quarks (upper two lines)
 and $d$ quarks (lower lines). This lowest moment of $h_1$ is found
 from the $n=1$ case of Eq.~(\ref{h1_op}). We have made two 
 determinations of $\delta q$, once from the expectation value of
 the operator $O^{2,4}$ calculated with a stationary nucleon 
 (squares), 
 and once
 from the operator $O^{1,2}$ measured for a nucleon with one unit of 
 momentum in the 1-direction 
(circles).
 (In both cases our nucleon spin is polarised
 in the 2-direction.) If Lorentz symmetry has been restored on the lattice, 
 both determinations would agree, which they do within the errors, though
 for $O^{1,2}$ the errors are large. In the heavy quark limit
 the $u$ and $d$ quark contributions to $\delta q$ are $+4/3$ and $-1/3$
 respectively. We see that the $u$ contribution has dropped below this
 value for our quark masses. 
  
  It is interesting to compare the lattice results with the results
 of a recent QCD sum rule calculation~\cite{qcd_sum}, which finds
 $\delta u = 1.33 \pm 0.53$ and $\delta d = 0.04 \pm 0.02$ at the
 scale $1 \rm{GeV}^2$. The small value of $\delta d$ differs from the
 results of lattice gauge theory, (see Fig.~\ref{h1} or~\cite{aoki}). 

  As mentioned in the previous section,
 the comparison between $h_1$ and $g_1$ gives us an impression
 of how relativistic the quarks in a nucleon are. Comparing 
 $\delta q$ and the valence quark contribution to 
 $\Delta q \equiv \int^1_0 d x g_1(x)$, as reported in~\cite{stephenson},  
 we see that they 
 are rather similar, (a conclusion also reached in~\cite{aoki,MIT})
 showing that for the
 quark masses used, which are approximately in the strange quark range, 
 a non-relativistic description of the spin structure is reasonable.

\section*{ACKNOWLEDGMENTS}
\label{acknowledgements}
 We wish to thank DESY-Zeuthen, where the 
 numerical calculations were performed on an APE
(Quadrics QH2).

\end{document}